\documentclass[
    ,final            
  ]
  {aipproc}

\layoutstyle{6x9}

\def\la{\mathrel{\hbox{\rlap{\hbox{\lower4pt\hbox{$\sim$}}}\hbox{$<$}}}}

\begin{document}

\title{Rapid Oscillations in Cataclysmic Variables, and a Comparison 
with X-Ray Binaries}

\author{Brian Warner and Patrick A. Woudt}{
  address={Department of Astronomy, University of Cape Town, Rondebosch 7700, 
South Africa}
}

\begin{abstract}
      We compare some of the properties of rapid oscillations in cataclysmic 
variables and X-Ray binaries. In addition to the earlier recognition that both 
types possess the same correlation between high and low frequency 
quasi-periodic oscillations, we have now found that the dwarf nova VW Hyi in 
its late stages of outburst shows the 1:2:3 oscillation harmonics that are 
seen in some neutron star and black holes X-Ray binaries.
  We point out that the behaviour of the dwarf nova WZ Sge has some 
similarities to those of accreting millisecond pulsars.
\end{abstract}

\maketitle


\section{Introduction}

    The rapid oscillations seen in cataclysmic variables (CVs) and X-Ray 
binaries (XRBs), although separated by orders of magnitude in time scales, 
show many common features in their phenomenology. The former have 
been recognised since the singular discovery of highly stable 71 sec 
modulations in the optical light curve of DQ Her, exactly 50 years ago, in 
July 1954 (Walker 1956), followed by the opening of a floodgate of 
phenomena associated with the discovery of similar but less stable 
oscillations in outbursting dwarf novae (Warner \& Robinson 1972). The 
first observed XRB rapid modulations date to the days of EXOSAT and the 
quasi-periodic 5 -- 50 Hz flux variations seen in the low mass XRB GX 5-1 
(van der Klis et al.~1985). With the advent of RXTE in 1996 the time 
resolution was increased to where kilohertz oscillations could be detected, 
with the result that now a quarter of the neutron star XRBs (Swank 2004) 
and a sixth of the black hole XRBs (McClintock \& Remillard 2004) have 
been observed with quasi-periodic oscillations (QPOs).

\section{Oscillations in CVs}

   The rich phenomenology of oscillations in CVs has been reviewed 
recently by Warner (2004). There are at least three distinct types of rapid 
oscillation in CVs, which can exist separately or simultaneously:

\begin{enumerate}
\item{Dwarf Nova Oscillations (DNOs).   These are oscillations typically in 
the range 3 -- 40 s, usually appearing in high mass transfer ($\dot{M}$) discs, i.e. 
dwarf novae in outburst and nova-like variables. They have not been 
observed in all such CVs, and are of varying amplitude, often disappearing 
altogether. They show a period-$\dot{M}$ relationship, with shortest periods at 
highest $\dot{M}$, and are of moderate coherence with small abrupt changes of 
period and almost continuous phase noise. `Double DNOs' are occasionally 
seen, with frequency differences equal to the frequency of the QPO (see (3) 
below) present at the time. DNOs are sinusoidal modulations; the 
companion in a double DNO is at longer period and usually possesses a first 
harmonic (Warner \& Woudt 2002, hereafter WW1; Woudt \& Warner 2002, 
hereafter WW2).}
\item{Longer-period DNOs.  These lpDNOs have periods $\sim$ 4 times those of 
DNOs and show little if any variation with $\dot{M}$ (or optical luminosity). 
They are less commonly present than DNOs, are occasionally seen even in 
quiescence of dwarf novae, and can also appear doubled on rare occasions 
(Warner, Woudt \& Pretorius 2003, hereafter WWP).}
\item{Quasi-Periodic Oscillations.  The QPOs of interest here are those with 
periods $P_{QPO} \sim 15 P_{DNO}$ (there are others of longer period probably related to 
the rotation of the slower rotating primaries - Patterson et al.~2002). These 
QPOs have a short coherence length, typically growing and decaying in $\sim$ 5 
cycles, and changing phase or period on that time scale if they are longer 
lived (WW2).}
\end{enumerate}

    It is useful to have a model in mind that ties these variations together. The 
following is the one advanced by WW1, which in its fundamentals was first 
proposed by Paczynski (1978). This is, in principle, an extension of the 
standard intermediate polar model (see Chapters 7 and 8 of Warner (1995)) 
to lower field strengths. The recent discovery of kilogauss magnetic fields 
in a number of DA stars (Aznar Cuadrado et al.~2004) demonstrates that 
such low fields are probably common in isolated white dwarfs and therefore 
also in CVs (though they are below the detection limit of direct 
measurement).

    Provided that the primary has a sufficiently weak magnetic field 
($\la 5 \times 10^5$ G) accretion through a disc onto its surface generates a freely 
moving equatorial belt. There is spectroscopic evidence of such a belt after 
dwarf nova outbursts (e.g. Sion et al.~1996; Godon et al.~2004). If the 
primary's field is stronger then the system is an intermediate polar with very 
stable rotation. The rapidly rotating belt will enhance whatever field the 
primary has, resulting in magnetic channeling of accreting mass even for 
weak field primaries, but onto the equatorial belt. The low inertia of the belt 
is what allows it to be spun up and down by magnetic connection to the 
inner edge of the accretion disc -- giving the $\dot{M}$-$P_{DNO}$ relationship seen 
in outbursting dwarf novae. We refer to this as the Low Inertia Magnetic 
Accretor (LIMA) model (WW1).

    There is a key observation that supports the idea that at least some of the 
white dwarfs in dwarf novae have magnetic fields strong enough to channel 
accretion, despite the absence yet of direct detection by spectroscopic or 
polarimetric means, and this comes from the X-Ray eclipse of OY Car 
observed shortly after outburst. Wheatley \& West (2003) deduce that the X-Ray 
emitting region is considerably smaller than the white dwarf and is at 
high primocentric latitude.

     The QPOs are thought (by WW1) to be due to a prograde traveling wave 
at the inner edge of the disc, probably excited by field winding arising from 
the non-synchronous rotation of the disc and the primary. The traveling 
wave can intercept and/or reprocess radiation from near the primary, 
thereby generating the optical QPOs. The revolving anisotropic radiation 
from the accretion zones on the belt (that causes the DNOs) can sweep 
across the traveling wave and be reprocessed at the beat period, producing 
the double DNOs.

  The lpDNOs may be an additional channel of accretion from the disc, but 
along field lines that connect to the body of the primary, rather than to the 
belt.

   Not all CVs in their high $\dot{M}$ phases show DNOs or QPOs. This shows 
that there is a parameter that determines the presence or absence of 
oscillations. In the LIMA model this is simply the strength of the magnetic 
field of the equatorial belt, which is determined by both the field of the 
primary and any shear enhancement that takes place.

\section{The Evolution of DNOs through Outburst in VW Hydri}

   At maximum luminosity of a VW Hyi outburst DNOs are seen at 14 s, 
both in soft X-Rays (van der Woerd et al.~1987) and the optical (WW2). 
Until almost the end of outburst DNOs are seen only intermittently, but 
when visible they are found to increase slowly and systematically in period 
to $\sim$ 20 s when the star has descended to $\sim$ 1.5 mag above minimum. At this 
phase of outburst DNOs are almost always present; as the star continues to 
fade WW2 discovered that there is a rapid increase in period, doubling to $\sim$ 40 s 
in about 5 h. This has been ascribed (WW1) to propellering, caused 
when the equatorial belt with its associated magnetic field is rotating at 
higher angular velocity than the inner edge of the accretion disc, and is 
consistent with the near cessation of EUV flux (which is a direct monitor of 
$\dot{M}$ onto the primary) precisely during the phase of rapid DNO 
deceleration (Mauche 2002). Following this deceleration phase there has 
been uncertainty in the evolution of the DNOs. Figure 8 of WW2 showed 
that the DNO periods at some point decrease by a factor of about two and 
the oscillations themselves become apparently less coherent. We have 
gathered more light curves of VW Hyi covering these final phases of 
outburst, the analysis of which has uncovered a remarkable behaviour.

\begin{figure}
  \includegraphics[height=.55\textheight]{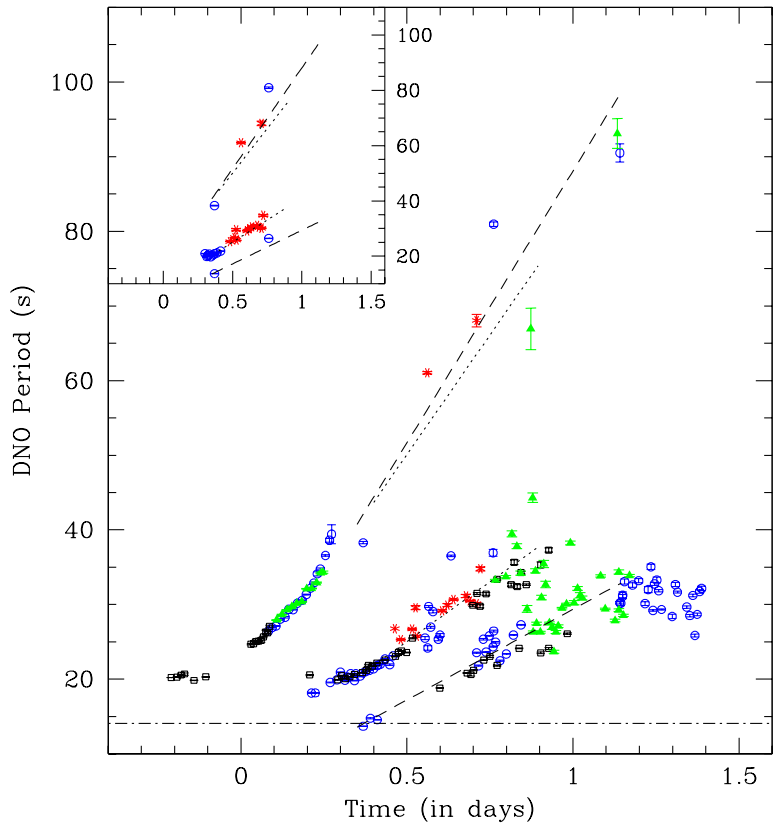}
  \caption{The evolution of DNO periods at the end of normal and super outbursts of the
dwarf nova VW Hyi. Different symbols mark the various kind of outbursts, ranging from
short (red asterisks), normal (blue open circles), long (black open squares) to super (green 
filled triangles) outbursts. The dotted and dashed lines show the results of a least-squares fit to the
first and second harmonic, respectively, and are multiplied by a factor of two and three
to illustrate the inferred evolution of the fundamental DNO period. The dashed-dotted horizontal
line at 14.1 s represents the minimum observed DNO period at maximum brightness. The inset highlights two
observing runs in which the fundamental, the first and/or second harmonic of the DNO period were
present simultaneously.}
  \label{warnerf1}
\end{figure}

\subsection{Late stage DNOs in VW Hydri}

When the DNOs have increased to $P_{DNO} \sim 39$ s there is a sudden frequency 
doubling -- seen in Figure\ref{warnerf1}. We have not managed to be at the telescope, 
with a clear sky, when this happens, but our various observing runs suggest 
that it happens in less than $\sim$ 15 min, and (as seen by the later behaviour) 
may happen essentially instantaneously. As we do not have the crucial 
information we cannot yet claim that the period exactly halves, but again 
from subsequent behaviour we can be fairly certain that the fundamental has 
been replaced by its first harmonic. The period of the now dominant 1st 
harmonic continues to increase systematically and with little scatter as VW 
Hyi decreases in brightness until at a period of $\sim$ 28 s a new periodicity 
appears, which within error of measurement is at a period 2/3rds of the 1st 
harmonic, so here are sure that we are dealing with a 2nd harmonic.

\begin{figure}
  \includegraphics[height=.63\textheight]{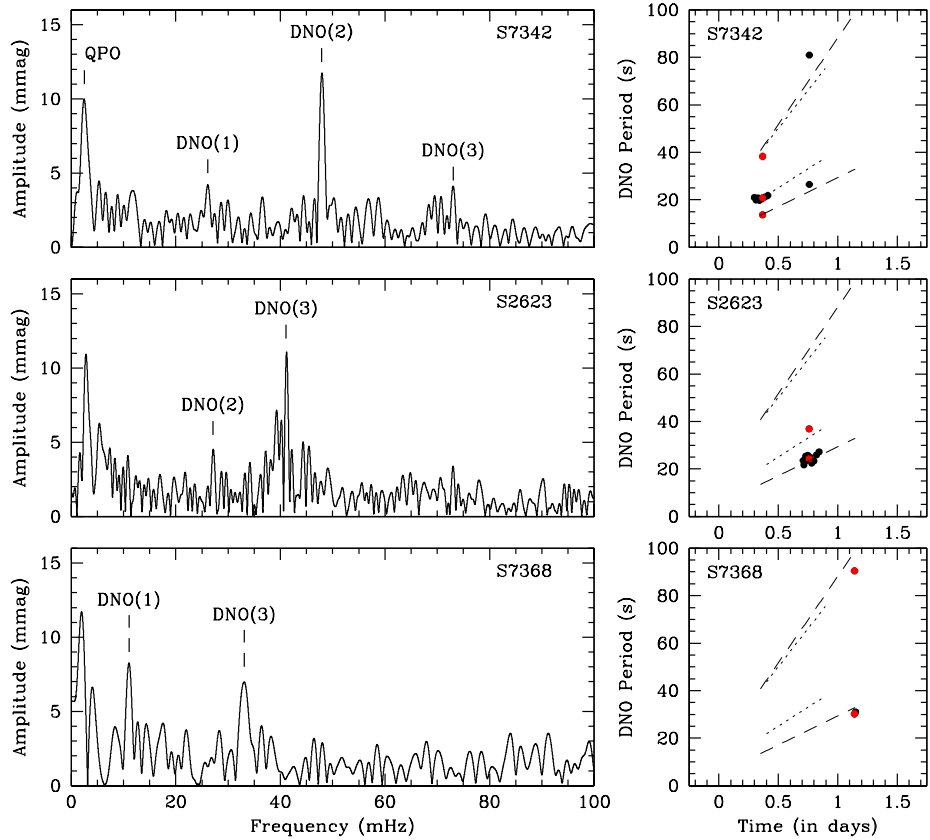}
  \caption{Examples of Fourier transforms of VW Hyi in which combinations of the 
fundamental and the first two harmonics of the DNO period are present.}
  \label{warnerf2}
\end{figure}

   The evolution of the DNOs from this point is complex in the sense that 
there is a gradual move from dominance by the 1st harmonic to dominance 
by the 2nd harmonic, with often both being simultaneously present -- and, 
very importantly, there is an occasional appearance of the associated 
fundamental, showing that the description as fundamental, 1st and 2nd 
harmonics is quantitatively justified.  Figure~\ref{warnerf2} shows examples of Fourier 
transforms (FTs) with various combinations of the harmonics present.
 
\begin{figure}
  \includegraphics[height=.55\textheight]{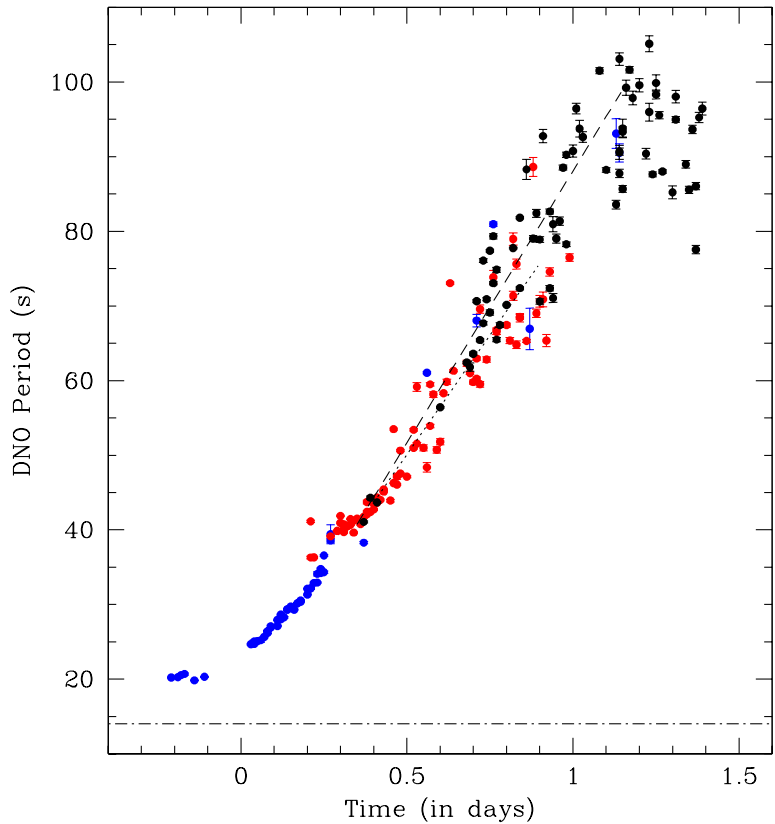}
  \caption{The evolution of the fundamental DNO period in VW Hyi at the end of normal
and super outbursts. The different colours indicate the observed fundamental (blue), first
harmonic (red) -- multiplied by two --, and second harmonic (black) -- multiplied by three.
The dotted, dashed and dashed-dotted lines are as in Figure~\ref{warnerf1}.}
  \label{warnerf3}
\end{figure}

   In Figure~\ref{warnerf3} we show a modified form of Figure~\ref{warnerf1}, in which the harmonics 
have been replaced by their implied fundamentals. From this we see that the 
fundamental continues to increase at a rapid rate, with increasing scatter, 
until it reaches $\sim$ 105 s, after which VW Hyi has reached quiescence and we 
have not detected DNOs. The fundamental period increases almost linearly 
from 30 s to 105 s in $\sim$ 27.5 h, i.e. $\dot{P} = 1.06 \times 10^{-3}$. 

   We have not detected such 1:2:3 ratios in the DNOs of any CV other than 
VW Hyi -- but VW Hyi is also unique in having DNOs that are most 
prominent at the end of its outburst.

\subsection{Similarity to XRBs}

\begin{figure}
  \includegraphics[height=.55\textheight]{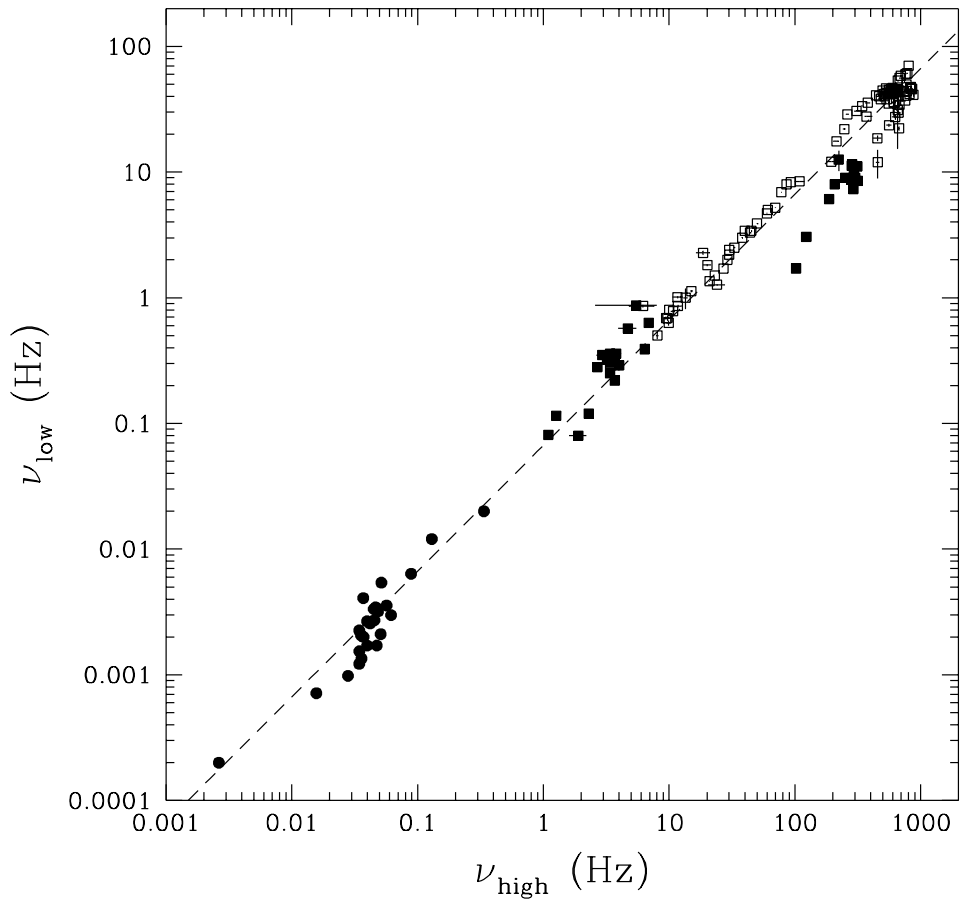}
  \caption{The Two-QPO diagram for X-ray binaries (filled squares: black hole binaries,
open squares: neutron star binaries) and 26 CVs (filled circles). Each CV is plotted
only once. The X-Ray binary data are from Belloni et al.~(2002) and were kindly provided by 
T.~Belloni. The dashed line marks $P_{QPO}/P_{DNO} = 15$ (From WWP).}
  \label{warnerf4}
\end{figure}

   As pointed out in WW1 and WWP, CVs and XRBs show the same 
correlation of their high frequency and low frequency oscillations, viz. the 
ratio of $\sim$ 15. This relationship, which extends over 6 orders of magnitude in 
frequency, is shown in Figure~\ref{warnerf4}.

    The 3:2:1 period ratios seen in VW Hyi are similar to those already 
known for a few years in XRBs (McClintock \& Remillard 2004). For 
example, for black hole binaries, XTE J1550-564 has strong X-Ray signals 
at 276 and 184 Hz and a weak signal at 92 Hz, which are in the ratio 3:2:1; 
GRO J1655-40 has 450 and 300 Hz oscillations (Remillard et al.~2002); and 
240 and 160 Hz (ratio 3:2) are present in H1743-322 (Homan et al.~2004). 
For the neutron star XRB, Sco X-1, Abramowicz et al.~(2003) claim a 3:2 
period ratio.

\section{A Model for Harmonics in VW Hyi}

    A possible explanation of the appearance of harmonics in the DNOs of 
VW Hyi is as follows. At high $\dot{M}$, i.e. maximum of outburst, the DNOs 
are rarely present, but when they are they are seen (in different outbursts) at 
14.1 s. We identify this as the minimum period of the equatorial belt -- i.e. 
the keplerian period at the surface of the primary, which leads to a mass of 
0.70 M$_{\odot}$. At high $\dot{M}$ the inner edge of the disc is close to the primary and 
we can see only the upper accretion zone once per rotation. But, at an 
orbital inclination of $\sim 63^{\circ}$ for VW Hyi, the inner edge of the disc 
eventually retreats (with lowering $\dot{M}$) sufficiently for us to see the zone 
on the far side (i.e., no longer hidden by the primary); when this happens the 
keplerian period at the inner edge (which will be similar to the rotation 
period of the magnetically coupled equatorial belt), with the above 
parameters, should be $\sim$ 45 s. This accounts for the observed frequency 
doubling at $P_{DNO} \sim 40$ s.

   As the inner parts of the disc are cleared out by the expanding 
magnetosphere, eventually the observed DNO should become the 
reprocessed beam -- the traveling (QPO generating) wave subtends the 
largest and nearest reprocessing site. The frequency of the observed DNO 
would then be the first harmonic of the difference between the frequencies 
of the equatorial belt and the traveling wave, which we can call $\tilde{\omega}$.

   The traveling wave (according to WW1) is a region of deceleration and 
pile-up of disc material, and therefore constitutes a region of higher density 
at the inner edge of the disc. The magnetic field of the equatorial belt 
sweeps across this at frequency $\tilde{\omega}$, thereby modulating $\dot{M}$ onto the 
primary. This is an application of the beat-frequency model, first introduced 
for CVs (Warner 1983) and later independently for XRBs (Alpar \& Shaham 
1985; Lamb et al.~1985). The outcome is that the first harmonic, at frequency 
2$\tilde{\omega}$ will have sidebands at $\pm$ $\tilde{\omega}$, i.e. the amplitude 
modulated frequency set $\tilde{\omega}$, 2$\tilde{\omega}$ and 3$\tilde{\omega}$. The modulations 
at $\tilde{\omega}$ and 2$\tilde{\omega}$ will have additional components, 
of other amplitudes and phases, so their relative amplitudes will not be that 
of simple amplitude modulation, and may vary with time. Similar effects are 
seen in the orbital sidebands of intermediate polars (Warner 1986).

   It should be noted that in this model there is no physical oscillation at the 
second harmonic, 3$\tilde{\omega}$, it is purely a Fourier component in the light curve, as 
actually observed.
   
\section{The Robertson-Leiter Model}

  As black holes are not supposed to have magnetic fields it is not 
automatically obvious that the same magnetically channeled accretion 
model for CVs and neutron stars is applicable to black hole XRBs. 
However, in a series of papers Robertson \& Leiter have suggested that all of 
the central objects in galactic black hole candidate XRBs have magnetic 
moments. The evidence comes from (1) the conclusion that in neutron star 
binaries the power-law part of the X-Ray flux spectrum arises in the neutron 
star magnetosphere, where calculations and observations are in close 
agreement, and the recognition that the same model works for black hole 
XRBs (Robertson \& Leiter 2002); (2) the realisation that a magnetic star 
collapsing towards its event horizon can be slowed by the radiation pressure 
ensuing from pair annihilation that itself is a result of pair production by the 
compressed magnetic field -- such a ``Magnetospheric, Eternally Collapsing 
Object'' maintains an intrinsic magnetic moment outside its event horizon 
for orders of magnitude longer than a Hubble time (Robertson \& Leiter 
2003); (3) models of jets that arise from thin disc interaction with MECOs 
produce the Radio-IR correlation with Mass and X-Ray luminosity observed 
in neutron star and Black Hole binaries (and also in AGNs) (Robertson \& 
Leiter 2004). 

  This recognition that black holes can have magnetospheres that interact 
with accretion discs permits a unified model that may explain the similarity 
of behaviour of DNOs and QPOs in CVs and X-Ray binaries seen in Fig. 4.

\section{WZ Sagittae}

    In quiescence the dwarf nova WZ Sge has optical modulations at 27.87 
and 28.95 s; their beat period at 744 s is commonly seen in the light curve 
and is responsible for recurring absorption dips. These have been 
interpreted in terms of the LIMA model (WW1), where the magnetic field 
of the primary is strong enough to prevent easy slippage of the equatorial 
belt in quiescence (Warner 2004), in which case the 27.87 s period should 
be thought of as an lpDNO.

    During outburst weak oscillations near 6.5 s have been detected (Knigge 
et al.~2002), which may be the DNOs associated with the lpDNOs (i.e. with 
rotation of white dwarf itself, and also oscillations at 15 s that increased to 
18 s (Welsh et al.~2003), which could be from a spun-up equatorial belt.

   The reason for including WZ Sge here is because it appears to be an 
analogue of the accreting millisecond pulsars (MSPs) in X-Ray binaries, 
where $\dot{M}$ is very low, the spin rate is relatively high (a factor of only a 
few above rotational break-up period), and only a small amount of gas 
manages to leak onto the primary (Galloway et al.~2002; Rappaport, Fregeau 
\& Spruit 2004).  X-Ray transients containing MSPs have disc instabilities 
just as in dwarf novae. $\dot{M}$ during outburst is directly measured from the 
X-Ray flux; the millisecond flux pulsations are sinusoidal and are present 
during the whole factor of 50 -- 100 of $\dot{M}$ variation during outburst. 
Optical and UV DNOs are present during superoutbursts of WZ Sge, and in 
quiescence, but their behaviour is complicated by reprocessing of the high 
energy rotating beam off various stationary and moving components in the 
binary system. The magnetic field of the primary in WZ Sge appears not to 
be strong enough to prevent a freely moving equatorial band from 
developing during the highest $\dot{M}$ delivered in outburst. In contrast, in the 
MSP XRBs the pulses are monoperiodic and have small but measurable 
period derivatives, but here there is no free equatorial belt and we observe 
only the X-Rays from the accretion region itself.

\section{Conclusions}

   CVs show an increasing range of rapid modulation phenomena that have 
analogues among the XRBs. It appears increasingly likely that magnetically 
channeled accretion is the underlying cause for these properties, implying 
magnetic moments for some stellar mass black holes.




\begin{theacknowledgments}
BW is supported by research funds from the University of Cape Town; PAW 
is supported by research funds from the University, from the National 
Research Foundation, and by a strategic award given by the University to BW.
\end{theacknowledgments}

\bibliographystyle{aipprocl} 


\end{document}